\documentclass[prd,twocolumn,showpacs,preprintnumbers,amsmath,amssymb,superscriptaddress,floatfix,nofootinbib]{revtex4}

\usepackage{graphicx}
\usepackage{amsmath}
\usepackage{amsfonts}
\usepackage{amssymb}
\usepackage{color}
\usepackage{multirow}
\usepackage[colorlinks, citecolor=blue,anchorcolor=red,menucolor=red, linkcolor=red,filecolor=red,runcolor=red,urlcolor=blue,frenchlinks=red]{hyperref}

\begin{document}

\title{Search for the $D^*\bar{D}^*$ molecular state $Z_c(4000)$ in the reaction $B^{-} \rightarrow J/\psi \rho^0 K^{-}$}

\author{Yang Zhang}
\affiliation{School of Physics and Microelectronics, Zhengzhou University, Zhengzhou, Henan 450001, China}

\author{En Wang} \email{wangen@zzu.edu.cn}
\affiliation{School of Physics and Microelectronics, Zhengzhou University, Zhengzhou, Henan 450001, China}

\author{De-Min Li}
\affiliation{School of Physics and Microelectronics, Zhengzhou University, Zhengzhou, Henan 450001, China}

\author{Yu-Xiao Li}
\affiliation{School of Physics and Microelectronics, Zhengzhou University, Zhengzhou, Henan 450001, China}

\begin{abstract}
Based on the prediction of a $D^*\bar{D}^*$ molecular state $Z_c(4000)$ with isospin $I=1$ in the coupled channel approach, we suggest to search for this state in the reaction  $B^- \to J/\psi \rho^0 K^-$.  By taking into account the final state interactions of $J/\psi \rho$ and $D^{*0}\bar{D}^{*0}$, and the contribution from the $K_1(1270)$ resonance, we find that the $J/\psi\rho$ mass distribution shows a peak around 4000~MeV, which could be associated to the $D^*\bar{D}^*$ molecular state $Z_c(4000)$.
 Searching for the $Z_c(4000)$ in the reaction $B^- \to J/\psi \rho^0 K^-$ is crucial to understand the internal structures of the exotic hadrons,  and our predictions can be tested by the Belle II and LHCb in future.
\end{abstract}

\maketitle

\section{Introduction}
\label{sec:introduction}
In the last decades, a lot of charmonium-like states, named as $X$, $Y$, $Z$ states, were discovered experimentally,
which provides a good platform to study the multiquark dynamics~\cite{Brambilla:2019esw,Olsen:2012zz,Chen:2016qju,Oset:2016lyh,Guo:2017jvc,Lebed:2016hpi,Olsen:2017bmm}. Among various explanations of the internal structure of these states, hadronic molecule, analogous to the deuteron, plays an important role since the predictions of those states can be made with controlled uncertainty~\cite{Oset:2016lyh,Guo:2017jvc}.

Generally speaking, it is not easy to identify one state as the hadronic molecular state dynamically generated from hadron-hadron interactions, since there exists the possible mixing of various configurations. One way to unambiguously identify a hadronic molecule or multiquark state is the observation of resonances decaying into a heavy quarkonium plus a meson with nonzero isospin meson, or plus a light baryon.
For instance, the first charged charmonium-like state, $Z_c(4430)$ was reported in the $\pi^- \psi(2S)$ mass distribution of the $B\to K \pi^- \psi(2S)$ by the Belle Collaboration~\cite{Choi:2007wga,Chilikin:2013tch}, and confirmed by the LHCb Collaboration seven years later~\cite{Aaij:2014jqa}. In 2013, the $Z_c(3900)$ was observed in the $\pi^- J/\psi$ invariant mass distribution of the $e^+e^-\to \pi^+\pi^- J/\psi$ by the BESIII and Belle Collaborations~\cite{Ablikim:2013mio,Liu:2013dau}. By now several $Z_c$ states were reported experimentally in different processes~\cite{Brambilla:2019esw}, and the hadronic molecules and tetraquark states are proposed for their internal structures, which opens a new window for understanding the non-perturbative properties of QCD.

Searching for more $Z_c$ states, especially around the lowest-lying thresholds $D\bar{D}$, $D\bar{D}^*$, and $D^*\bar{D}^*$,  would be helpful to understand the internal structures of $Z_c$ states, and also the hadron spectroscopy. Recently, one resonance $Z_c(4000)$, with $D^*\bar{D}^*$  molecule nature and quantum numbers of $I^G(J^{PC})=1^-(2^{++})$, was predicted in Ref.~\cite{Aceti:2014kja}, where a thorough investigation of the $D^*\bar{D}^*$ and $J/\psi\rho$ interactions was performed by considering the vector exchanges within the local hidden gauge approach. The channel $J/\psi\rho$ is open for the decay, and is responsible for a width of the order of 100~MeV.  Due to the quark components and the isospins of $J/\psi$ and $\rho$ in the final state, any resonance observed in the $J/\psi\rho$ channel would be unambiguously interpreted as an exotic state $Z_c$, rather than the $c\bar{c}$ state. A $Z_c$ state with mass around 4000~MeV and $J^P=2^+$ was also predicted in the QCD sum rules~\cite{Qiao:2013dda,Wang:2014gwa,Khemchandani:2013iwa} and the color flux-tube model~\cite{Deng:2014gqa}.

The weak decays of  heavy mesons and baryons turn out to be an important tool to identify molecular~\cite{Oset:2016lyh,Chen:2016qju,Lebed:2016hpi,Olsen:2017bmm,Oset:2016nvf, Lu:2016roh,Wang:2015pcn,Chen:2015sxa}. One example is that the analysis of the LHCb measurements about the reaction $B\to J/\psi\phi K$~\cite{Aaij:2016nsc} shows the existence of the $X(4160)$ resonance with the $D^*_s\bar{D}^*_s$ molecular nature~\cite{Wang:2017mrt}, and also provides a natural interpretation of the quite large width of the $X(4140)$~\cite{Aaij:2016nsc}. In addition, two $D^*\bar{D^*}$  molecular states, $X(3930)$ and $X(3940)$, predicted in the coupled channel approach, where the vector-vector interactions are described by the Lagrangian of the hidden gauge formalism~\cite{Molina:2009ct}, are also found to play an important role in the $J/\psi\omega$ mass distribution of the reaction $B^+\to J/\psi\omega K$~\cite{Dai:2018nmw}.
In this paper, we will investigate the role of the $Z_c(4000)$ in the reaction $B^{-} \rightarrow J/\psi \rho^0 K^{-}$. So far, only
 the Belle Collaboration has reported the observation of the exclusive decay process $B^+\to J/\psi K_1(1270)^+, K_1(1270)\to K\pi\pi$, and measured the branching fraction of Br$[B^+\to J/\psi K^+_1(1270)]=(1.80\pm0.34\pm0.39)\times 10^{-3}$~\cite{Abe:2001wa}.
It also shows that the clustering near $M_{\pi\pi}\approx M_\rho$ and $M_{K\pi\pi}\approx 1.27$~GeV is consistent with expectations for $K_1(1270)\to K\rho$ decays~\cite{Abe:2001wa}.
Since the dominant decay channel of the $K_1(1270)$ is $\rho K$~\cite{PDG2018}, it implies that the reaction  $ B^-\to J/\psi \rho^0 K^-$ is accessible experimentally.

It should be pointed out that the $X(3872)$ was observed in the decay $B^\pm \to J/\psi K^\pm \pi^+\pi^-$ by the Belle, BaBar, CDF, and LHCb Collaborations~\cite{Choi:2003ue,Choi:2011fc,Aubert:2008gu,Abulencia:2005zc,Aaij:2013zoa,Aaij:2015eva}. However, there has been no significant structure around 4000~MeV in the $J/\psi\pi^+\pi^-$ mass distribution of the $B^\pm\to J/\psi K^\pm\pi^+\pi^-$~\cite{Choi:2003ue,Choi:2011fc,Aubert:2008gu,Abulencia:2005zc,Aaij:2013zoa,Aaij:2015eva}, which implies that the branching fraction of $B^\pm\to Z_c(4000)K^\pm\to J/\psi \pi^+\pi^- K^\pm$, with $K^\pm$ in $D$-wave, is less than the one of $B^\pm\to X(3872)K^\pm\to J/\psi \pi^+\pi^- K^\pm$, with $K^\pm$ in $P$-wave.
For the $J/\psi\pi^+\pi^-$ mass distribution, one of the dominant background sources comes from the $K_1(1270)$, which mainly contributes to the region of $4300<M_{J/\psi\rho}<4700$~MeV if the events of $\rho$ meson are selected, and we will discuss this issue later. In this paper, we will show that the more precise measurement of the $J/\psi \pi^+\pi^-$ mass distribution around 4000 MeV, and the better understanding of the background, are much important for checking the existence of the predicted $Z_c(4000)$ state.

This paper is organized as follows. In Sec.~\ref{sec:formalism}, we will present the mechanism of the reaction $ B^-\to J/\psi \rho^0 K^-$, the results and the discussions are shown in Sec.~\ref{sec:results}. Finally, the summary is given in Sec.~\ref{sec:summary}.

\section{FORMALISM}
\label{sec:formalism}
\begin{figure}[h]
\begin{center}
\includegraphics[width=0.5\textwidth]{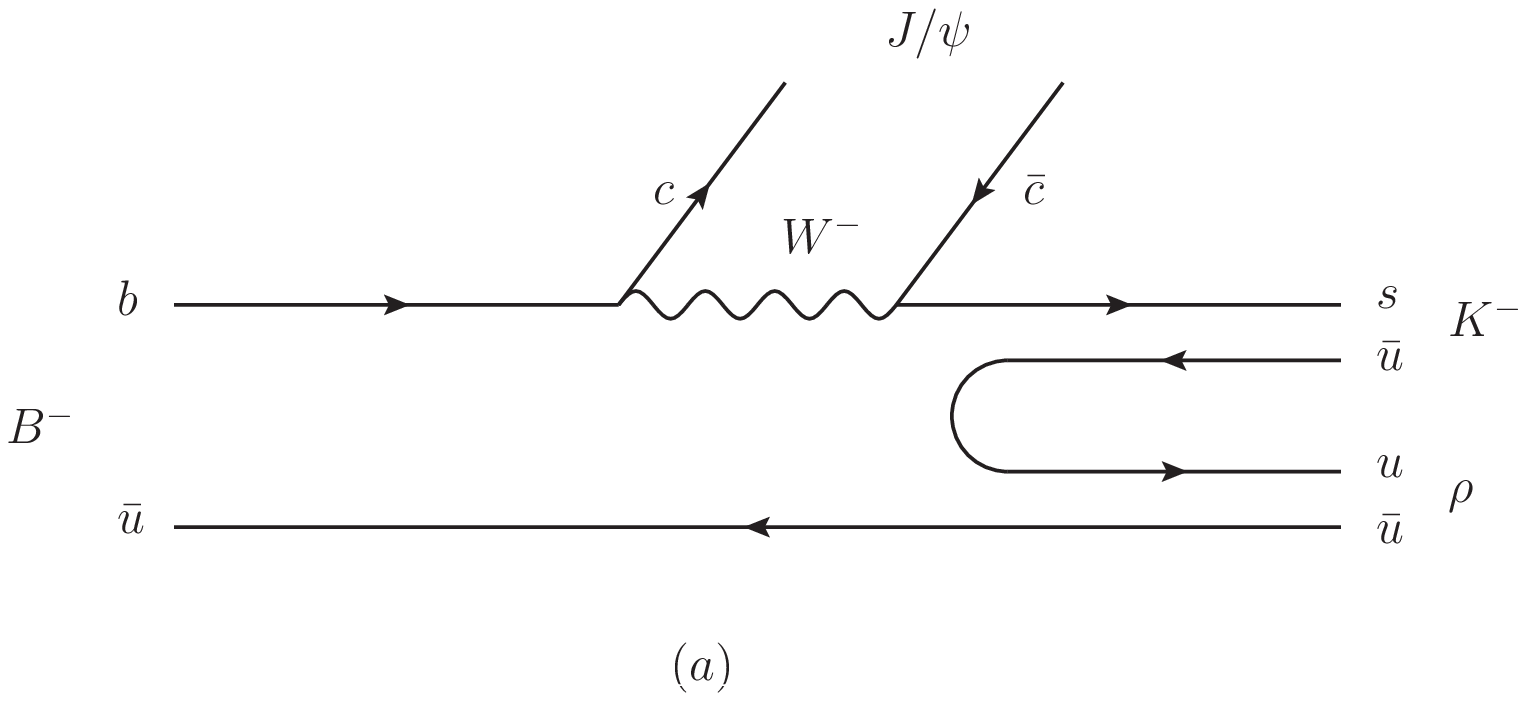}
\includegraphics[width=0.5\textwidth]{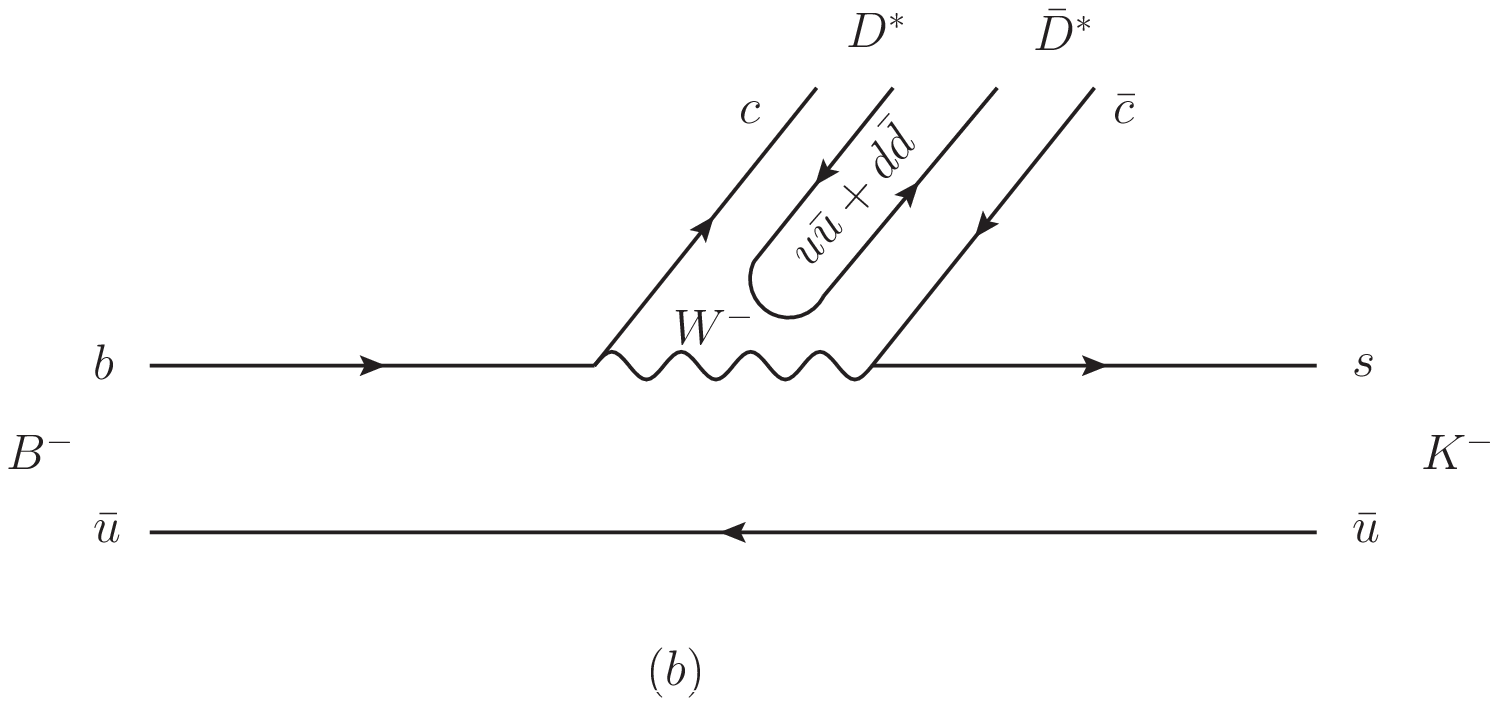}
\includegraphics[width=0.5\textwidth]{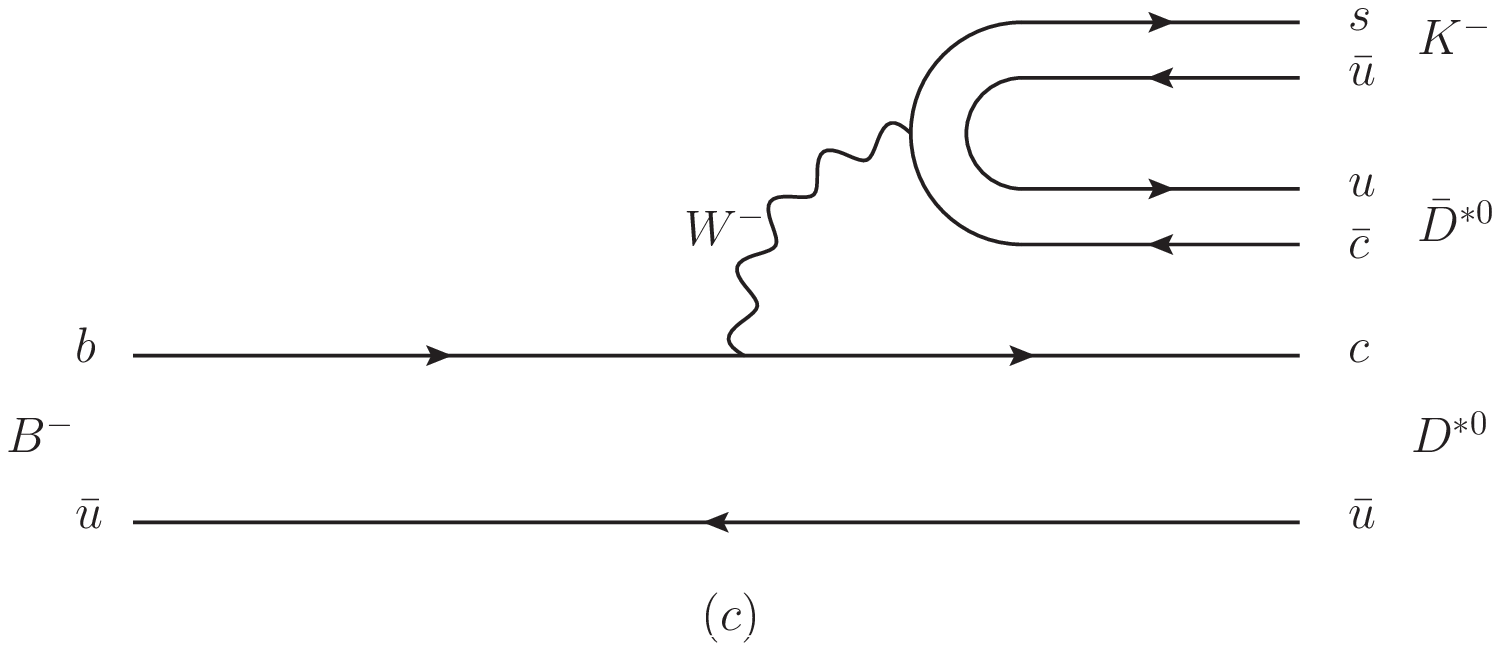}
\caption{The microscopic quark level production of the $B^-$ decay. (a) The internal emission of the $B^-\to J/\psi s\bar{u}$ decay and hadronization  of the $s\bar{u}$ through $\bar{u}u$ with vacuum quantum numbers. (b) The internal emission of the $B^-\to K^- c\bar{c}$ decay and hadronization  of the $c\bar{c}$ through $\bar{q}q$ with vacuum quantum numbers. (c) The external emission of the $B^-\to D^{*0} \bar{c}s$ decay and hadronization  of the $\bar{c}s$ through $\bar{q}q$ with vacuum quantum numbers. }
\label{fig:quarklevel}
\end{center}
\end{figure}

\begin{figure}[h]
\begin{center}
\includegraphics[width=0.235\textwidth]{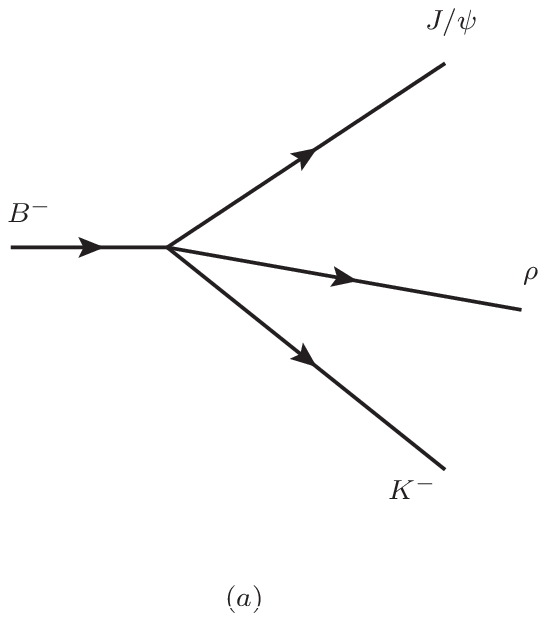}
\includegraphics[width=0.235\textwidth]{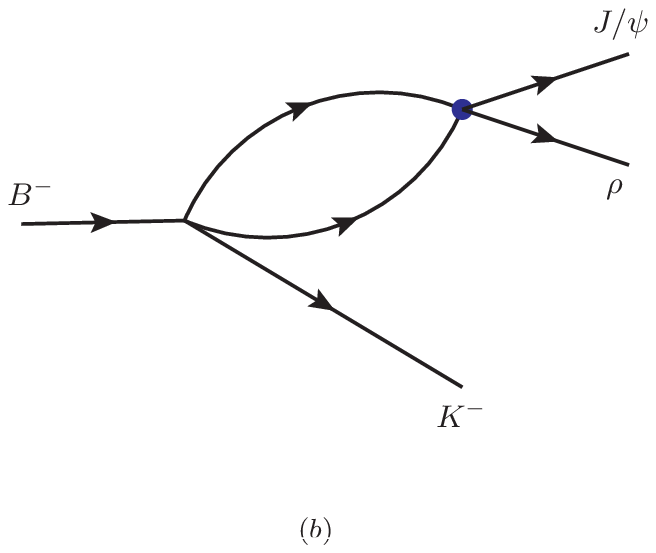}
\includegraphics[width=0.24\textwidth]{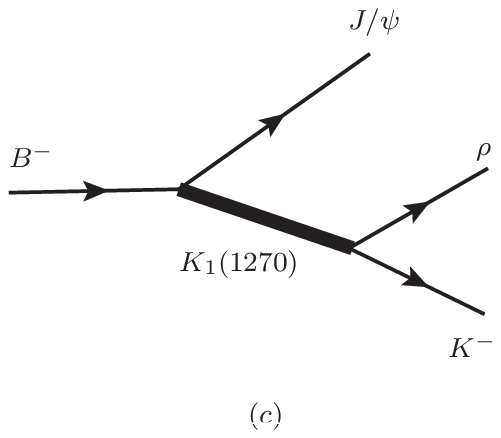}
\caption{The mechanisms for the $B^-\to J/\psi\rho^0 K^-$ reaction. (a) the tree diagram, (b) the $J/\psi\rho$ final state interaction, (c) the term of the intermediate $K_1(1270)$. }
\label{fig:feynman}
\end{center}
\end{figure}

In analogy to the Refs.~\cite{Dai:2018nmw,Wang:2017mrt}, the mechanism of the reaction $B^-\to J/\psi \rho^0 K^-$ at the quark level can be depicted in Fig.~\ref{fig:quarklevel}. The $b$ quark first weakly decays into a $c$ quark and a $W^-$ boson, and then the $W^-$ boson couples to a $\bar{c}$ quark and an $s$ quark. Fig.~\ref{fig:quarklevel}(a) shows the internal emission, where the $c$ and $\bar{c}$ go into $J/\psi$, and the $s\bar{u}$ component is hadronized with $\bar{u}u$ pair, created from the vacuum with the quantum numbers of vacuum, to $\rho K^-$.  Since the state $Z_c(4000)$ couples strongly to the $D^*\bar{D}^*$, the $D^*\bar{D}^*$ system can be produced primarily, followed by the transition to the final state $J/\psi\rho$. Figure~\ref{fig:quarklevel}(b) shows the internal emission mechanism of the reaction $B^-\to D^*\bar{D}^* K^-$, where the $c$ and $\bar{c}$ hadronize with the $\bar{q}q$ pair, created from the vacuum, to the final state $D^*\bar{D}^*$. Because the isospin of the created $\bar{q}q$ is 0, which leads to the isospin $I=0$ for the $D^*\bar{D}^*$ system, the diagram of Fig.~\ref{fig:quarklevel}(b) has no contribution to the reaction of $B^-\to J/\psi \rho^0 K^-$. In addition, we also have the mechanism of external emission as shown in Fig.~\ref{fig:quarklevel}(c), which is color-favored with respect to the internal emission. Here the $s\bar{c}$ component from the $W^-$ decay, together with the $\bar{u}u$, is hadronized to produce the $\bar{D}^{*0} K^-$, and the remaining $c\bar{u}$ leads to the $D^{*0}$.

The tree level diagrams of the $B^-\to J/\psi \rho^0 K^-$ reaction, and the final state interactions of $J/\psi \rho$ and $D^{*0}\bar{D}^{*0}$, are shown in Figs.~\ref{fig:feynman}(a) and (b), respectively.
The tree level amplitude for the $B^-\to J/\psi \rho^0 K^-$ decay in $S$-wave can be expressed as,
\begin{equation}
\mathcal{M}^{(a)}=A \times  \vec\epsilon_{J/\psi}\cdot \vec\epsilon_{\rho},\label{eq:tree}
\end{equation}
where the $\vec\epsilon_{J/\psi}$ and $\vec\epsilon_\rho$ are the polarization vectors for the $J/\psi$ and $\rho$, respectively, and $A$ stands for the normalization factor of the vertex $B^-\to J/\psi\rho^0 K^-$.
Note that we work on the rest frame of the resonance produced, where the  momenta of the $J/\psi$ and $\rho$ are small with respect to their masses, thus we neglect the $\epsilon^0$ component. This is actually very accurate for these momenta as can be seen in Appendix A of Ref.~\cite{Sakai:2017hpg}.
For the final state interactions of the $J/\psi\rho$ and $D^*\bar{D}^*$ final state interaction as shown in Fig.~\ref{fig:feynman}(b),
we need the $K^-$ in $D$-wave to match the angular momentum of  the $B^-$,
the amplitude is given by~\cite{Dai:2018nmw,Wang:2017mrt},
\begin{eqnarray}
\mathcal{M}^{(b)}&=& \frac{B}{|\vec{k}_{\rm ave}|^2} \left( G_{J/\psi \rho} t_{J/\psi\rho, J/\psi\rho} \right. \nonumber \\
 && \left.  +3C\,\frac{1}{\sqrt{2}} G_{D^*\bar{D}^*}  t^{I=1}_{D^*\bar{D}^*, J/\psi\rho}  \right)\nonumber \\
&&\times \left(\vec{\epsilon}_{J/\psi}\cdot \vec{k}\,\vec{\epsilon}_{\rho}\cdot \vec{k} -\frac{1}{3}|\vec{k}|^2 \vec{\epsilon}_{J/\psi}\cdot \vec{\epsilon}_\rho  \right), \label{eq:4000}
\end{eqnarray}
where $\vec{k}$ is the momentum of the $K^-$ in the $J/\psi\rho$ rest frame, and we  include a factor $1/|\vec{k}_{\rm ave}|^2$, with $|\vec{k}_{\rm ave}|=1000$~MeV, in order to make the strength $B$ with the same dimension as $A$. The factor $1/\sqrt{2}$ is the Clebsch-Gordan coefficient for the $D^{*0}\bar{D}^{*0}$ system with isospin $I=1$. In order to explicitly consider the factor 3 relative to the enhancement of the external emission mechanism of Fig.~\ref{fig:quarklevel}(c), we
write $3C$ for the weight of the mechanism relative to $D^{*0} \bar{D}^{*0}$ primary production. We will vary the value
of $C$ around unity, but we can anticipate that it hardly
changes the shape of the distribution obtained.

\begin{figure}[h]
\begin{center}
\includegraphics[width=0.5\textwidth]{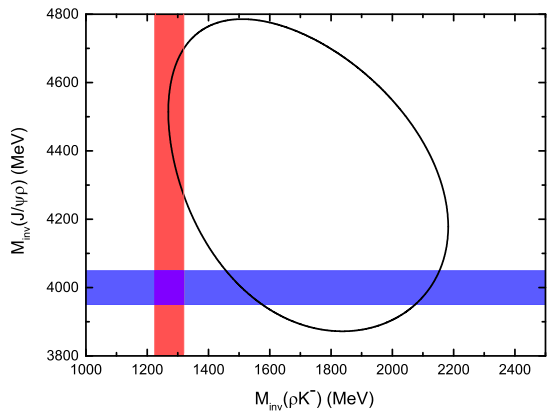}
\caption{The Dalitz plot of the $B^-\rightarrow J/\psi\rho K^-$ reaction. The bands colored by blue and red correspond to the energy regions ($M-\Gamma/2,M+\Gamma/2$) of the $Z_c(4000)$ and $K_1(1270)$ resonances, respectively. Here we take $M_{K_1}=1272$~MeV and $\Gamma_{K_1}=90$~MeV for the $K_1(1270)$ from the PDG~\cite{PDG2018},  and $M_{Z_c}=4000$~MeV and $\Gamma_{Z_c}=100$~MeV for the $Z_c(4000)$ from Ref.~\cite{Aceti:2014kja}.}
\label{fig:dalitz}
\end{center}
\end{figure}

The $ G_{J/\psi \rho}$ and $G_{D^*\bar{D}^*}$ are the loop functions, and we use the dimensional regularization as,
\begin{eqnarray}
G_{i}=& \frac{1}{16 \pi^2} \left\{ \alpha_i + \ln
\frac{m_1^2}{\mu^2} + \frac{m_2^2-m_1^2 + s}{2s} \ln
\frac{m_2^2}{m_1^2} \right. \nonumber\\
&+ \frac{p}{\sqrt{s}} \left[ \ln(s-(m_2^2-m_1^2)+2
p\sqrt{s}) \right. \nonumber \\
&+ \ln(s+(m_2^2-m_1^2)+2 p\sqrt{s})  \nonumber  \\
&
 - \ln(-s+(m_2^2-m_1^2)+2 p\sqrt{s}) \nonumber \\
&\left.\left.- \ln(-s-(m_2^2-m_1^2)+2 p\sqrt{s}) \right]\right\},
\label{eq:loopfunction}
\end{eqnarray}
where the subtraction constants $\alpha_1=-2.3$ and $\alpha_2=-2.6$ ($i=1,2$ corresponding to the channels of $D^*\bar{D}^*$ and $J/\psi\rho$), $\mu=1000$~MeV, same as Ref.~\cite{Aceti:2014kja}. $p$ is the three-momentum of the mesons $D^*$ or $J/\psi$ in the rest frame of $D^*\bar{D}^*$ or $J/\psi\rho$, respectively,
\begin{equation}
p=\frac{\sqrt{(s-(m_1+m_2)^2)(s-(m_1-m_2)^2)}}{2\sqrt{s}},
\end{equation}
with $m_{1,2}$ being the masses of the mesons in the $i$th channel.

The transition amplitudes of $t_{J/\psi\rho, J/\psi\rho}$ and $t^{I=1}_{D^*\bar{D}^*, J/\psi\rho}$ are taken by solving the Bethe-Salpeter equation, as shown in  Eq.~(8) of Ref.~\cite{Aceti:2014kja}.

In addition, the $K^-\rho$ can also undergo the final state interaction. Ref.~\cite{Abe:2001wa} has observed the $B\to J/\psi K_1(1270) $ with Br$[B^+\to J/\psi K^+_1(1270)]=(1.80\pm 0.34\pm 0.39) \times 10^{-3} $, and no evidences of other high-mass kaons are seen.
Since the dominant decay channel of the $K_1(1270)$ is $\rho K$ (Br$[K_1(1270)\to \rho K]=(42\pm6)\%$~\cite{PDG2018}), we expect the resonance $K_1(1270)$ will play an important role in the $\rho K^-$ invariant mass distribution, as shown in Fig.~\ref{fig:feynman}(c), and the contributions from the other high-mass kaons could be safely neglected. Although some theoretical studies show that the $K_1(1270)$ has a two-pole structure~\cite{Geng:2006yb,Wang:2019mph,Wang:2020pyy}, the contribution from the $K_1(1270)$ will not affect the peak structure of the $Z_c(4000)$ in the $J/\psi\rho$ invariant mass distribution, according to the Dalitz diagram of the  $B^-\to J/\psi \rho^0 K^-$ of Fig.~\ref{fig:dalitz}.
For simplicity, we will include the amplitude for the $K_1(1270)$ contribution with a Breit-Wigner form,
\begin{equation}\label{eq:K1270}
\mathcal{M}^{(c)}= \frac{ A'\times M^2_{K_1} \times \epsilon_{J/\psi}\cdot \epsilon_\rho}{M^2_{\rm inv}(K\rho)-M^2_{K_1}+ i M_{K_1} \Gamma_{K_1}},
\end{equation}
with $M_{K_1}=1272$~MeV, and $\Gamma_{K_1}=90$~MeV~\cite{PDG2018}. Now, we can write the full amplitude for the $B^-\to J/\psi\rho^0 K^-$ reaction,
\begin{eqnarray}
\mathcal{M}&=& \mathcal{M}^{(a)}+\mathcal{M}^{(b)}+ \mathcal{M}^{(c)}\nonumber \\
&=& A \times \, \vec\epsilon_{J/\psi}\cdot \vec\epsilon_\rho \times \nonumber \\
&& \left[1+\frac{\beta M^2_{K_1} }{M^2_{\rm inv}(K\rho)-M^2_{K_1}+ i M_{K_1} \Gamma_{K_1}} \right] \nonumber \\
&& +\frac{B}{|\vec{k}_{\rm ave}|^2} \left( G_{J/\psi \rho} t_{J/\psi\rho, J/\psi\rho} + \frac{3C}{\sqrt{2}}  G_{D^*\bar{D}^*} t^{I=1}_{D^*\bar{D}^*, J/\psi\rho}  \right)\nonumber \\
&&\times \left(\vec{\epsilon}_{J/\psi}\cdot \vec{k}\,\vec{\epsilon}_{\rho}\cdot \vec{k} -\frac{1}{3}|\vec{k}|^2 \vec{\epsilon}_{J/\psi}\cdot \vec{\epsilon}_\rho  \right) \nonumber \\
&=& A \times \, \vec\epsilon_{J/\psi}\cdot \vec\epsilon_\rho \times \left[t^{(a)}+t^{(c)} \right] \nonumber \\
&& +\frac{B}{|\vec{k}_{\rm ave}|^2} \left(\vec{\epsilon}_{J/\psi}\cdot \vec{k}\,\vec{\epsilon}_{\rho}\cdot \vec{k} -\frac{1}{3}|\vec{k}|^2 \vec{\epsilon}_{J/\psi}\cdot \vec{\epsilon}_\rho  \right) \times  t^{(b)},  \nonumber \\
&=& A \times \, \vec\epsilon_{J/\psi}\cdot \vec\epsilon_\rho \times t^{S} \nonumber \\
&& +\frac{B}{|\vec{k}_{\rm ave}|^2} \left(\vec{\epsilon}_{J/\psi}\cdot \vec{k}\,\vec{\epsilon}_{\rho}\cdot \vec{k} -\frac{1}{3}|\vec{k}|^2 \vec{\epsilon}_{J/\psi}\cdot \vec{\epsilon}_\rho  \right) \times  t^{D},\label{eq:amp_full}
\end{eqnarray}
where we define the terms from the $S$ and $D$ wave as,
\begin{eqnarray}
t^{S}&=&t^{(a)}+t^{(c)}\nonumber \\
&=&1+\frac{\beta M^2_{K_1} }{M^2_{\rm inv}(K\rho)-M^2_{K_1}+ i M_{K_1} \Gamma_{K_1}} \\
t^{D}&=& t^{(b)} \nonumber \\
&=&G_{J/\psi \rho} t_{J/\psi\rho, J/\psi\rho} + \frac{3C}{\sqrt{2}}  G_{D^*\bar{D}^*} t^{I=1}_{D^*\bar{D}^*, J/\psi\rho}, \nonumber \\ \label{eq:dwave}
\end{eqnarray}
with $\beta= A'/A$ standing for the relative weight of the contribution from the $K_1(1270)$ resonance.

With the above amplitudes, the mass distribution of the decay width is given as,
\begin{eqnarray}
\frac{d\Gamma^{2}}{d M^{2}_{J/\psi\rho} d M_{\rho K}^{2}}&=&\frac{1}{(2\pi)^{3}}\frac{1}{32M^{3}_{B^{-}}}\sum |\mathcal{M}|^{2} .
\end{eqnarray}
Since the $\vec{\epsilon}_{J/\psi}\cdot \vec{\epsilon}_\rho$ and $\left(\vec{\epsilon}_{J/\psi}\cdot \vec{k}\,\vec{\epsilon}_{\rho}\cdot \vec{k} -\frac{1}{3}|\vec{k}|^2 \vec{\epsilon}_{J/\psi}\cdot \vec{\epsilon}_\rho  \right)$ structures filter spin 0 and 2 respectively, they do not interfere when one sums over polarizations of all final states. Thus, the mass distribution can be rewritten by summing $\mathcal{M}$ over the final state polarizations,
\begin{eqnarray}
\frac{d\Gamma^{2}}{d M^{2}_{J/\psi\rho}d M_{\rho K}^{2}} &= & \frac{1}{(2\pi)^{3}}\frac{A^2}{32M^{3}_{B^{-}}}  \nonumber \\
&\times& \left(3  |t^{S}|^2+\frac{2B^2}{3A^2}\frac{ |\vec{k}|^4}{|\vec{k}_{\rm ave}|^4} |t^{D}|^2 \right) .\nonumber \\ \label{eq:dw_full}
\end{eqnarray}

\section{Results}
\label{sec:results}
\begin{figure}[h]
\begin{center}
\includegraphics[width=0.5\textwidth]{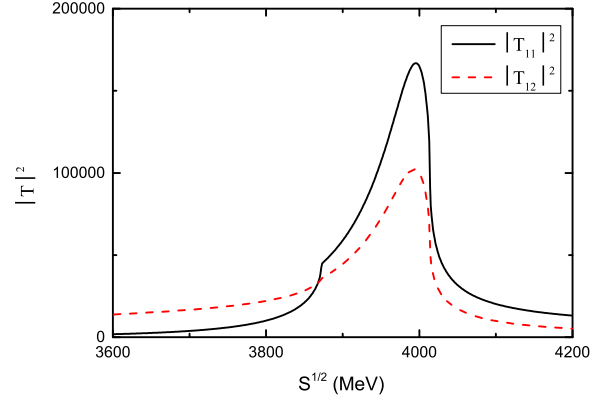}
\caption{Modulus squared of the transition amplitudes for $D^*\bar{D}^*\to D^*\bar{D}^*$ (the curve labeled as $|T_{11}|^2$) and $D^*\bar{D}^*\to J/\psi \rho $ (the curve labeled as $|T_{12}|^2$).}
\label{fig:tij}
\end{center}
\end{figure}

In this section, we will show our results with the above formalisms. First we present the modulus squared of the transition amplitudes $|T_{11}|^2$ for $D^*\bar{D}^*\to D^*\bar{D}^*$ and $|T_{12}|^2$ for $D^*\bar{D}^*\to J/\psi \rho $ in Fig.~\ref{fig:tij}, where one can see a peak around 4000~MeV, corresponding to the resonance $Z_c(4000)$ predicted in Ref.~\cite{Aceti:2014kja}.

\begin{figure}[h]
\begin{center}
\includegraphics[width=0.5\textwidth]{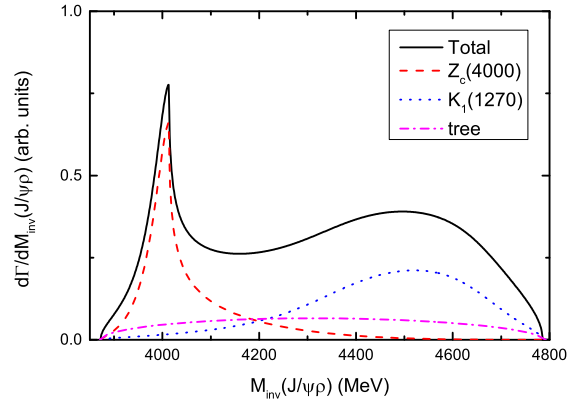}
\caption{The $J/\psi\rho$ mass distribution of the $B^-\rightarrow J/\psi\rho^0 K^-$ reaction. The curves labeled as `$Z_c(4000)$', `$K_1(1270)$', and `tree', correspond to the contributions of the $J/\psi\rho$ and $D^{*0}\bar{D}^{*0}$ final state interactions (Fig.~\ref{fig:feynman}(b)), the $K_1(1270)$ resonance (Fig.~\ref{fig:feynman}(c)), and the tree diagram (Fig.~\ref{fig:feynman}(a)), respectively. The `Total' curve shows the results of the full model.}
\label{fig:dw_jpsirho}
\end{center}
\end{figure}

\begin{figure}[h]
\begin{center}
\includegraphics[width=0.5\textwidth]{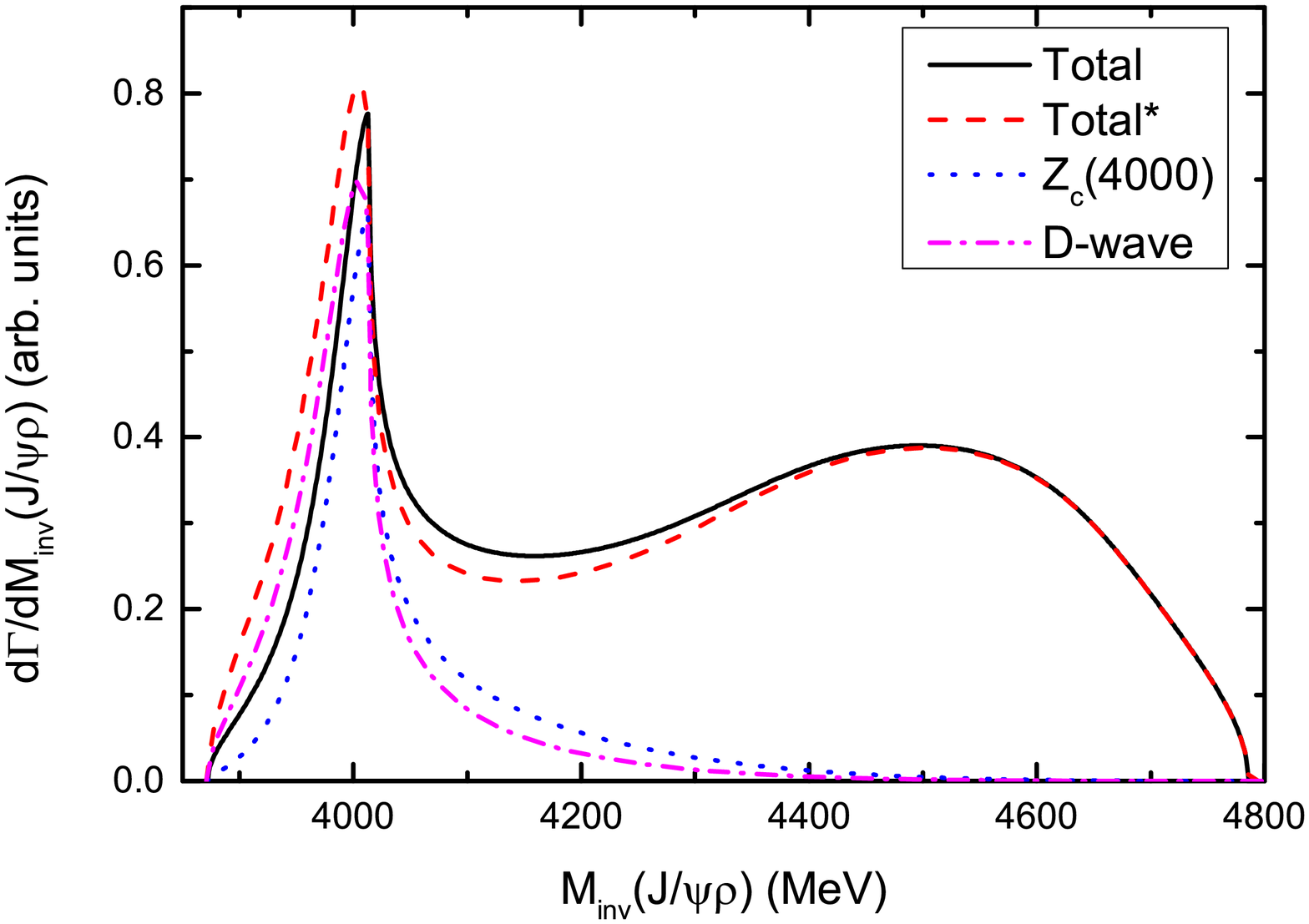}
\caption{The $J/\psi\rho$ mass distribution of the $B^-\rightarrow J/\psi\rho^0 K^-$ reaction. The curves labeled as `Total' and `$Z_c(4000)$' are the same as the ones of Fig.~\ref{fig:dw_jpsirho}, the curve labeled as `D-wave' shows the contributions from the tree diagram and the $J/\psi\rho$ and $D^{*0}\bar{D}^{*0}$ final state interactions, with $K^-$ in $D$-wave, and the `Total$^*$' curve is the total results by including the contribution of the tree diagram with $K^-$ in $D$-wave.}
\label{fig:dw_jpsirho_D}
\end{center}
\end{figure}

\begin{figure}[h]
\begin{center}
\includegraphics[width=0.5\textwidth]{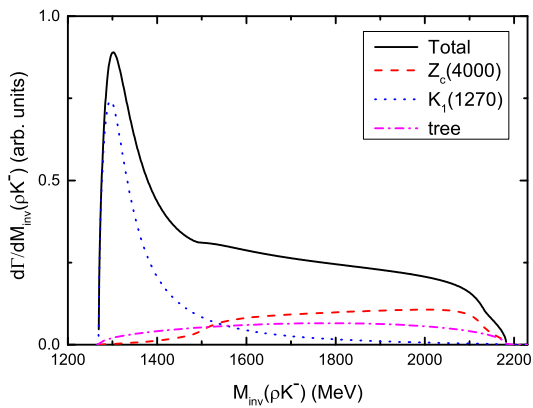}
\caption{The $\rho K^-$ mass distribution of the $B^-\rightarrow J/\psi\rho^0 K^-$ reaction. The explanations of the curves are same as those of Fig.~\ref{fig:dw_jpsirho}.}
\label{fig:dw_rhoK}
\end{center}
\end{figure}

\begin{figure}[h]
\begin{center}
\includegraphics[width=0.5\textwidth]{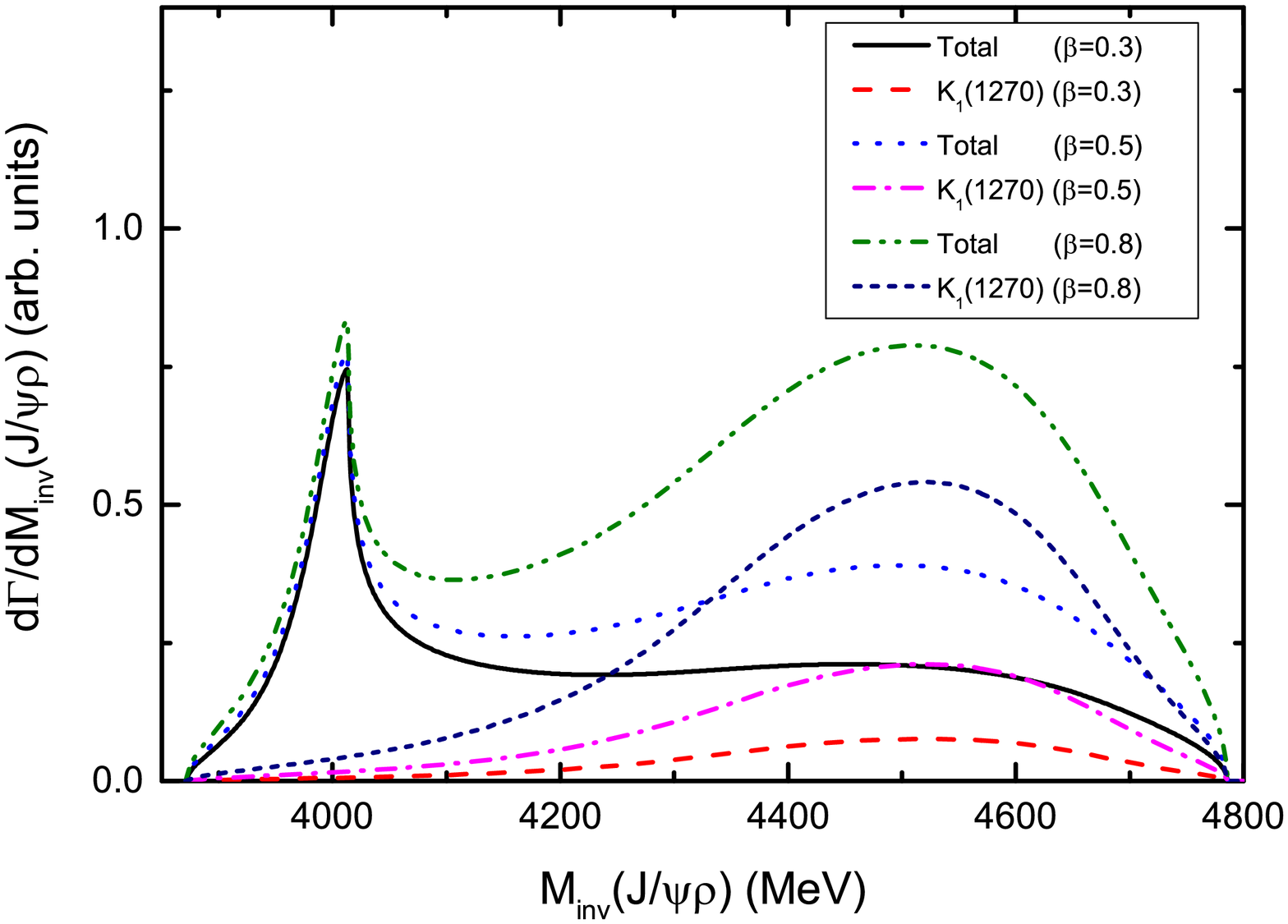}
\caption{The $J/\psi\rho$ mass distribution of the $B^-\rightarrow J/\psi\rho^0 K^-$ reaction for different values of $\beta$.}
\label{fig:dw_jpsirho_beta}
\end{center}
\end{figure}

\begin{figure}[h]
\begin{center}
\includegraphics[width=0.5\textwidth]{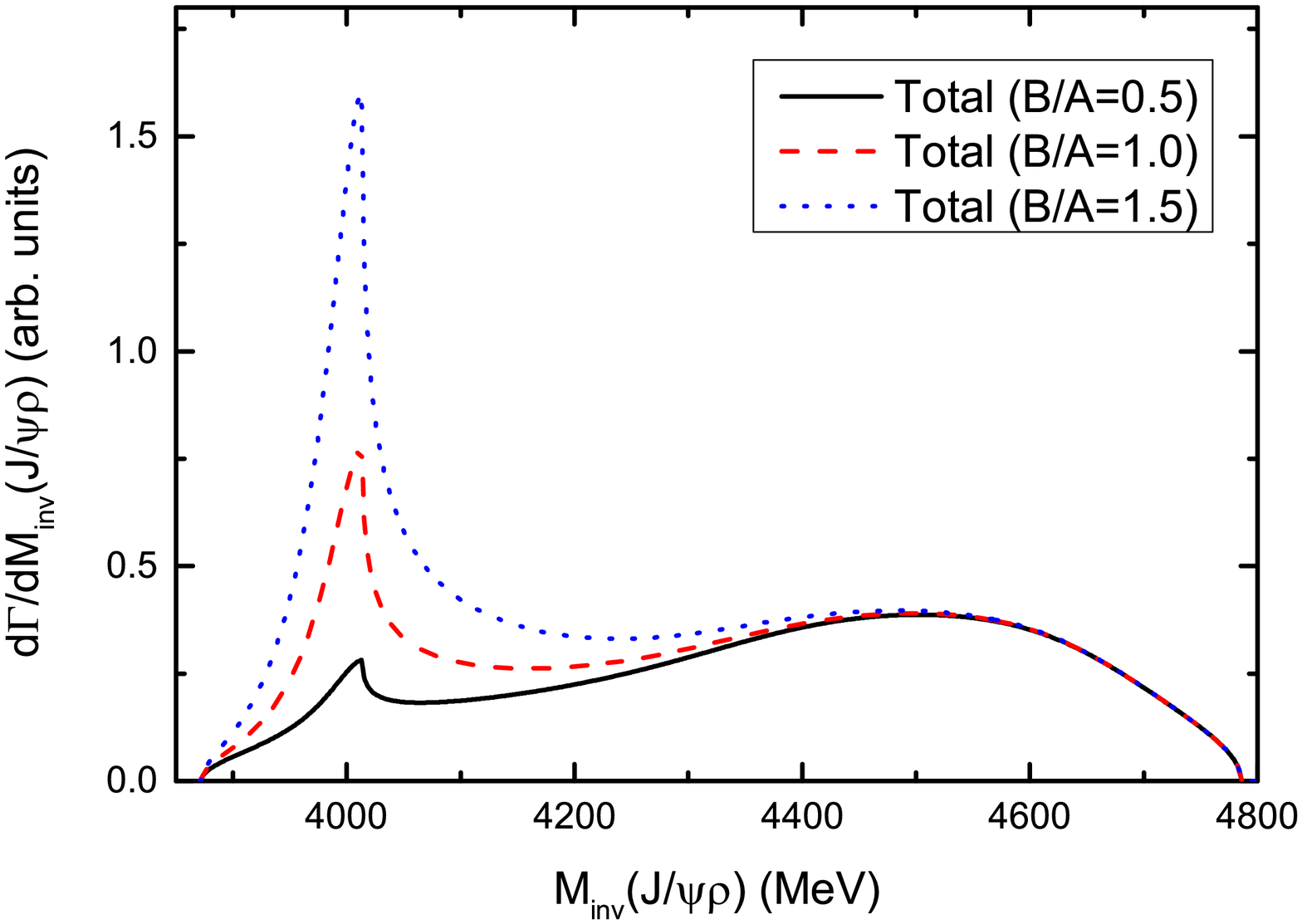}
\caption{The $J/\psi\rho$ mass distribution of the $B^-\rightarrow J/\psi\rho^0 K^-$ reaction for different values of $B$.}
\label{fig:dw_jpsirho_B}
\end{center}
\end{figure}

\begin{figure}[h]
\begin{center}
\includegraphics[width=0.5\textwidth]{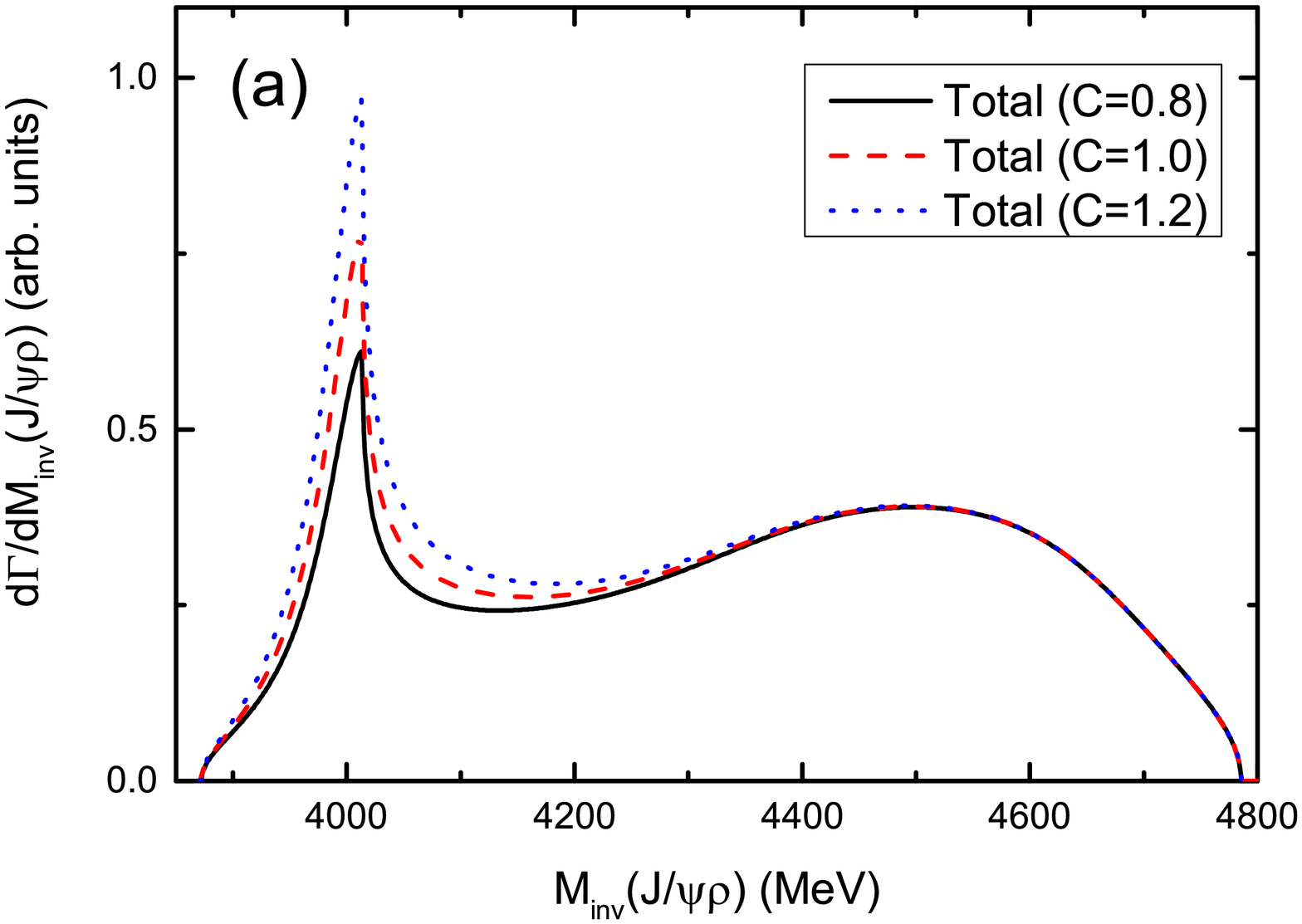}
\includegraphics[width=0.5\textwidth]{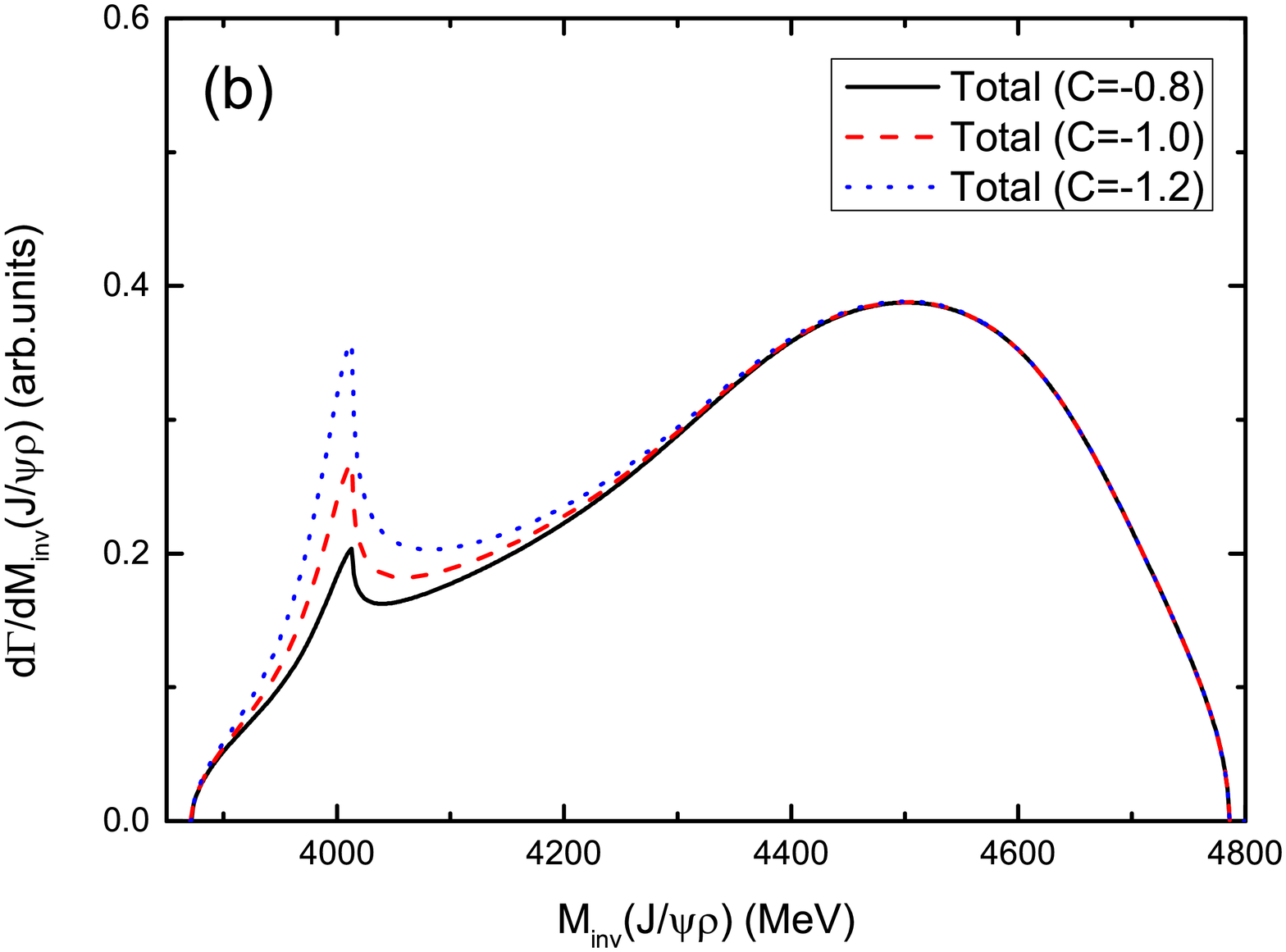}
\caption{The $J/\psi\rho$ mass distribution of the $B^-\rightarrow J/\psi\rho^0 K^-$ reaction, (a) for positive values of $C$, and (b) for negative values of $C$.}
\label{fig:dw_jpsirho_C}
\end{center}
\end{figure}

Before showing the mass distributions of the $B^-\to J/\psi\rho^0 K^-$ reaction, we need to choose the values of the free parameters of our model.
In addition to the arbitrary normalization $A$ of Eq.~(\ref{eq:dw_full}), we have three parameters, 1), $\beta$, the weight of the contribution from the $K_1(1270)$ resonance, 2), $B$, the weight of the contribution from the $J/\psi\rho$ and $D^{*0}\bar{D}^{*0}$ final state interactions, and 3), $C$, the weight of $D^{*0}\bar{D}^{*0}$ primary production, as shown in Eq.~(\ref{eq:amp_full}). We choose $\beta=0.5$ in order to give a sizable contribution from the $K_1(1270)$ resonance, and $C=1$.  Although we do not know the exact value of the $B/A$, one can  expect that  $B$ has a similar strength as $A$, since the primary production weight of the $J/\psi\rho$, shown in Fig.~\ref{fig:feynman}(b), is the same as that of the tree diagram  of Fig.~\ref{fig:feynman}(a),

Up to the arbitrary normalization $A$, we calculate the $J/\psi\rho$ and $\rho K^-$ mass distributions with $B/A=1$, as shown in Figs.~\ref{fig:dw_jpsirho} and \ref{fig:dw_rhoK}, respectively.
For the $J/\psi\rho$ mass distribution,
one can see a significant peak structure around 4000~MeV, which is associated to the $D^*\bar{D}^*$ molecular state $Z_c(4000)$.
The contributions from the tree diagram of Fig.~\ref{fig:feynman}(a) and the resonance $K_1(1270)$ have little effect on the peak position.
For the $\rho K^-$ mass distribution, Fig.~\ref{fig:dw_rhoK} shows a narrow peak close to the $\rho K^-$ threshold,  corresponding to the $K_1(1270)$ resonance, which is compatible with the $K\rho$ distribution reported by the Belle Collaboration~\cite{Abe:2001wa}.
Here we only consider the contribution from the tree diagram of Fig.~\ref{fig:feynman}(a) in $S$-wave, but the tree diagram with $K^-$ in $D$-wave also has contribution, which can be taken into account by replacing the $t^D=t^{\rm (b)}$ by $t^D=1++t^{\rm (b)}$ in Eq.~(\ref{eq:dwave}). In Fig.~\ref{fig:dw_jpsirho_D}, we can find the results including the contribution from the tree diagram with $K^-$ in $D$-wave is much small and can be safely neglected, by comparing the curve labeled as `Total$^*$' to the one of `Total'. For simplicity, we will neglect the contribution from the tree diagram with $K^-$ in $D$-wave in following calculations.

Next, we will show the $J/\psi\rho$ mass distributions by varying the values of the three parameters.
In Fig.~\ref{fig:dw_jpsirho_beta}, we present the $J/\psi\rho$ mass distributions with  $\beta=0.3,0.5,0.8$. From Fig.~\ref{fig:dw_jpsirho_beta}, we can conclude that contribution from the $K_1(1270)$ resonance does not modify the peak position of the $Z_c(4000)$ resonance markedly, and the peak structure is still clear even with a very large contribution from the $K_1(1270)$ resonance, because that the narrow peak structure of the $K_1(1270)$ almost does not contribute to the $J/\psi\rho$ mass distribution in the $3900\sim 4100$~MeV region, as shown in Fig.~\ref{fig:dalitz}.

The $J/\psi\rho$ mass distributions with the different values $B/A=0.5,1.0,1.5$ are shown in Fig.~\ref{fig:dw_jpsirho_B}. While the background contributions of the Figs.~\ref{fig:feynman}(a) and (c) become larger, the peak structure of the $Z_c(4000)$ will be weaker. Indeed, the ratio of $B/A$ can not  be determined with the present experimental information. Of course, whether one can find the signal of the $Z_c(4000)$ depends on the background, or the ratio of $B/A$. It should be pointed out that the weight of the tree diagram [Fig.~\ref{fig:feynman}(a)] is the same as the $J/\psi\rho$ final state interaction [Fig.\ref{fig:feynman}(b)], which implies that $B$ and $A$ should be in the same order of the magnitude if the contribution from the $K_1 (1270)$ is removed. Indeed, the $K_1 (1270)$ mainly contributes to the region of $M_{J/\psi\rho}>4200$~MeV, far away from the peak position of the $Z_c(4000)$, and the contribution from the $K_1 (1270)$ could be easily removed with a cut on the $\rho K^-$ invariant mass (for instance, remove the events of $M_{\rho K^-}<1400$~MeV). Thus, even if the $B/A$ is small, one can expect to find a peak around 4000~MeV with respect to the flat distribution from the background, by removing the contribution of the $K_1 (1270)$.

The parameter $3C$, corresponding to the relative weight of the external emission mechanism [Fig.~\ref{fig:quarklevel}(c)] with respect to the internal emission mechanism [Fig.~\ref{fig:quarklevel}(a)], should be around 3, since we take the number of the colors $N_c=3$. We  show the $J/\psi\rho$ mass distributions with $C=0.8,1.0,1.2$ in Fig.~\ref{fig:dw_jpsirho_C}(a). One can see that the signals of the $Z_c (4000)$ are always clear for the different values of $3C$ around 3. In addition, the $N_c$ scaling tell only the relative strength of the absolute values, and the relative sign between Fig.~\ref{fig:quarklevel}(a) and Fig.~\ref{fig:quarklevel}(c) is not fixed. Thus, we present the $J/\psi\rho$ mass distributions with $C=-0.8,-1.0,-1.2$ in Fig.~\ref{fig:dw_jpsirho_C}(b), where we can find the signal of the $Z_c(4000)$ is a little weaker, but still very clear.

\section{Summary}
\label{sec:summary}

In this work, we have studied the reaction of $B^-\to J/\psi\rho^0 K^-$, considering the $D^*\bar{D}^*$ molecular state $Z_c(4000)$ which couples to the $J/\psi\rho$ channel as well as the contribution from the $K_1(1270)$ resonance. The final state interactions of the $J/\psi\rho$ and $D^{*0}\bar{D}^{*0}$ with isospin $I=1$ are taken from the local hidden gauge approach.

Our results show that the $J/\psi\rho$ mass distribution has a peak structure, which can be associated to the $D^*\bar{D}^*$ molecular state $Z_c(4000)$. On the other hand, one can find a narrow peak structure close to the $\rho K^-$ threshold in the $\rho K^-$ mass distribution, which corresponds to the $K_1(1270)$ resonance. The contribution from the $K_1(1270)$ resonance does not affect the peak position of the $Z_c(4000)$.
As mentioned in the introduction, any resonance found in the $J/\psi\rho$ mass distribution would be unambiguously interpreted as an exotic state, therefore  we encourage our experimental colleagues to search for the $Z_c(4000)$ state in the reaction $B^-\to J/\psi\rho^0 K^-$.

\section*{Acknowledgements}

We warmly thank Eulogio Oset, Li-Sheng Geng, Ju-Jun Xie, and Feng-Kun Guo for useful discussions and comments.
This work is partly supported by the National Natural Science Foundation of China under Grant No. 11505158, the Key Research Projects of Henan Higher Education Institutions (No. 20A140027), and the Academic Improvement Project of Zhengzhou University.


\begin{thebibliography}{99}

\bibitem{Brambilla:2019esw}
  N.~Brambilla, S.~Eidelman, C.~Hanhart, A.~Nefediev, C.~P.~Shen, C.~E.~Thomas, A.~Vairo and C.~Z.~Yuan,
  The $XYZ$ states: experimental and theoretical status and perspectives,
  arXiv:1907.07583 [hep-ex].

\bibitem{Olsen:2012zz}
  S.~L.~Olsen,
  $X$, $Y$, $Z$ particles from Belle,
  Prog.\ Theor.\ Phys.\ Suppl.\  {\bf 193}, 38 (2012).

\bibitem{Chen:2016qju}
  H.~X.~Chen, W.~Chen, X.~Liu and S.~L.~Zhu,
  The hidden-charm pentaquark and tetraquark states,
  Phys.\ Rept.\  {\bf 639}, 1 (2016).

\bibitem{Oset:2016lyh}
  E.~Oset {\it et al.},
  Weak decays of heavy hadrons into dynamically generated resonances,
  Int.\ J.\ Mod.\ Phys.\ E {\bf 25}, 1630001 (2016).



\bibitem{Lebed:2016hpi}
R.~F.~Lebed, R.~E.~Mitchell and E.~S.~Swanson,
Heavy-Quark QCD Exotica,
Prog. Part. Nucl. Phys. \textbf{93} (2017), 143-194.


\bibitem{Olsen:2017bmm}
S.~L.~Olsen, T.~Skwarnicki and D.~Zieminska,
Nonstandard heavy mesons and baryons: Experimental evidence,
Rev. Mod. Phys. \textbf{90} (2018), 015003.

\bibitem{Guo:2017jvc}
  F.~K.~Guo, C.~Hanhart, U.~G.~Mei{\ss}ner, Q.~Wang, Q.~Zhao and B.~S.~Zou,
  Hadronic molecules,
  Rev.\ Mod.\ Phys.\  {\bf 90},  015004 (2018).

\bibitem{Choi:2007wga}
  S.~K.~Choi {\it et al.} [Belle Collaboration],
  Observation of a resonance-like structure in the $\pi^\pm \psi^\prime$ mass distribution in exclusive $B \to K \pi^\pm \psi^\prime$ decays,
  Phys.\ Rev.\ Lett.\  {\bf 100}, 142001 (2008).

\bibitem{Chilikin:2013tch}
  K.~Chilikin {\it et al.} [Belle Collaboration],
  Experimental constraints on the spin and parity of the $Z$(4430)$^+$,
  Phys.\ Rev.\ D {\bf 88},  074026 (2013).

\bibitem{Aaij:2014jqa}
  R.~Aaij {\it et al.} [LHCb Collaboration],
  Observation of the resonant character of the $Z(4430)^-$ state,
  Phys.\ Rev.\ Lett.\  {\bf 112}, 222002 (2014).

\bibitem{Ablikim:2013mio}
  M.~Ablikim {\it et al.} [BESIII Collaboration],
  Observation of a Charged Charmoniumlike Structure in $e^+e^-\to \pi^+ \pi^- J/\psi$ at $\sqrt{s}$ =4.26  GeV,
  Phys.\ Rev.\ Lett.\  {\bf 110}, 252001 (2013).

\bibitem{Liu:2013dau}
  Z.~Q.~Liu {\it et al.} [Belle Collaboration],
  Study of $e^+e^- \to \pi^+ \pi^- J/\psi$ and Observation of a Charged Charmoniumlike State at Belle,
  Phys.\ Rev.\ Lett.\  {\bf 110}, 252002 (2013).
  Erratum: [Phys.\ Rev.\ Lett.\  {\bf 111}, 019901 (2013)].

\bibitem{Aceti:2014kja}
  F.~Aceti, M.~Bayar, J.~M.~Dias and E.~Oset,
  Prediction of a $Z_c(4000)$ $D^* \bar D^*$ state and relationship to the claimed $Z_c(4025)$,
  Eur.\ Phys.\ J.\ A {\bf 50}, 103 (2014).

\bibitem{Qiao:2013dda}
  C.~F.~Qiao and L.~Tang,
  Interpretation of $Z_c(4025)$ as the hidden charm tetraquark states via QCD Sum Rules,
  Eur.\ Phys.\ J.\ C {\bf 74}, 2810 (2014).

\bibitem{Wang:2014gwa}
Z.~Wang,
Reanalysis of the $Y(3940)$, $Y(4140)$, $Z_c(4020)$, $Z_c(4025)$ and $Z_b(10650)$ as molecular states with QCD sum rules,
Eur.\ Phys.\ J.\ C \textbf{74} (2014), 2963.

\bibitem{Khemchandani:2013iwa}
  K.~P.~Khemchandani, A.~Martinez Torres, M.~Nielsen and F.~S.~Navarra,
  Relating $D^* \bar{D}^*$ currents with $J^P= 0^+,1^+$ and $2^+$ to $Z_c$ states,
  Phys.\ Rev.\ D {\bf 89},  014029 (2014).



\bibitem{Deng:2014gqa}
C.~Deng, J.~Ping and F.~Wang,
Interpreting $Z_c(3900)$ and $Z_c(4025)/Z_c(4020)$ as charged tetraquark states,
Phys.\ Rev.\ D \textbf{90} (2014), 054009.



\bibitem{Oset:2016nvf}
E.~Oset, H.~Chen, A.~Feijoo, L.~Geng, W.~Liang, D.~Li, J.~Lu, V.~K.~Magas, J.~Nieves, A.~Ramos, L.~Roca, E.~Wang and J.~Xie,
Study of reactions disclosing hidden charm pentaquarks with or without strangeness,
Nucl. Phys. A \textbf{954} (2016), 371-392.


\bibitem{Lu:2016roh}
J.~Lu, E.~Wang, J.~Xie, L.~Geng and E.~Oset,
The $\Lambda_{b}\rightarrow J/\psi K^{0}\Lambda$ reaction and a hidden-charm pentaquark state with strangeness,
Phys. Rev. D \textbf{93} (2016), 094009.

\bibitem{Wang:2015pcn}
E.~Wang, H.~Chen, L.~Geng, D.~Li and E.~Oset,
Hidden-charm pentaquark state in $\Lambda^0_b \to J/\psi p \pi^-$ decay,
Phys. Rev. D \textbf{93} (2016), 094001.

\bibitem{Chen:2015sxa}
H.~Chen, L.~Geng, W.~Liang, E.~Oset, E.~Wang and J.~Xie,
Looking for a hidden-charm pentaquark state with strangeness $S=−1$ from $\Xi_b^−$ decay into $J/\psi K^− \Lambda$,
Phys. Rev. C \textbf{93} (2016), 065203.

\bibitem{Aaij:2016nsc}
R.~Aaij \textit{et al.} [LHCb],
Amplitude analysis of $B^+\to J/\psi \phi K^+$ decays,
Phys.\ Rev.\ D \textbf{95} (2017), 012002.


\bibitem{Wang:2017mrt}
  E.~Wang, J.~J.~Xie, L.~S.~Geng and E.~Oset,
  Analysis of the $B^+\to J/\psi \phi K^+$ data at low $J/\psi \phi$ invariant masses and the $X(4140)$ and $X(4160)$ resonances,
  Phys.\ Rev.\ D {\bf 97}, 014017 (2018).


\bibitem{Molina:2009ct}
  R.~Molina and E.~Oset,
  The $Y(3940)$, $Z(3930)$ and the $X(4160)$ as dynamically generated resonances from the vector-vector interaction,
  Phys.\ Rev.\ D {\bf 80}, 114013 (2009).




\bibitem{Dai:2018nmw}
  L.~R.~Dai, G.~Y.~Wang, X.~Chen, E.~Wang, E.~Oset and D.~M.~Li,
  The $B^{+} \rightarrow J/\psi\omega K^{+}$ reaction and $D^{\ast} \bar{D}^{\ast}$ molecular states,
  Eur.\ Phys.\ J.\ A {\bf 55},  36 (2019).



\bibitem{Abe:2001wa}
  K.~Abe {\it et al.} [Belle Collaboration],
  Observation of $B\to J / \psi K_1(1270)$,
  Phys.\ Rev.\ Lett.\  {\bf 87}, 161601 (2001).

\bibitem{PDG2018}
  M.~Tanabashi {\it et al.} [Particle Data Group],
  Review of Particle Physics,
  Phys.\ Rev.\ D {\bf 98},  030001 (2018).





%
\bibitem{Choi:2003ue}
S.~Choi \textit{et al.} [Belle],
Observation of a narrow charmonium-like state in exclusive $B^\pm\to  K^\pm \pi^+ \pi^- J / \psi$ decays,
Phys. Rev. Lett. \textbf{91}, 262001 (2003).

\bibitem{Choi:2011fc}
S.~K.~Choi \textit{et al.} [Belle],
Bounds on the width, mass difference and other properties of $X(3872) \to \pi^+ \pi^- J/\psi$ decays,
Phys. Rev. D \textbf{84}, 052004 (2011).

\bibitem{Aubert:2008gu}
B.~Aubert \textit{et al.} [BaBar],
A Study of $B \to X(3872) K$, with $X{3872} \to J/\Psi \pi^{+} \pi^{-}$,
Phys. Rev. D \textbf{77}, 111101 (2008).

\bibitem{Abulencia:2005zc}
A.~Abulencia \textit{et al.} [CDF],
Measurement of the dipion mass spectrum in $X(3872) \to J/\psi \pi^+ \pi^-$ decays,
Phys. Rev. Lett. \textbf{96}, 102002 (2006).

\bibitem{Aaij:2013zoa}
R.~Aaij \textit{et al.} [LHCb],
Determination of the $X(3872)$ meson quantum numbers,
Phys. Rev. Lett. \textbf{110}, 222001 (2013).

\bibitem{Aaij:2015eva}
R.~Aaij \textit{et al.} [LHCb],
Quantum numbers of the $X(3872)$ state and orbital angular momentum in its $\rho^0 J\psi$ decay,
Phys. Rev. D \textbf{92}, 011102 (2015).



\bibitem{Sakai:2017hpg}
  S.~Sakai, E.~Oset and A.~Ramos,
  Triangle singularities in $B^-\rightarrow K^-\pi^-D_{s0}^+$ and $B^-\rightarrow K^-\pi^-D_{s1}^+$,
  Eur.\ Phys.\ J.\ A {\bf 54}, 10 (2018).


\bibitem{Geng:2006yb}
  L.~S.~Geng, E.~Oset, L.~Roca and J.~A.~Oller,
  Clues for the existence of two $K_1(1270)$ resonances,
  Phys.\ Rev.\ D {\bf 75}, 014017 (2007).


\bibitem{Wang:2019mph}
  G.~Y.~Wang, L.~Roca and E.~Oset,
  Discerning the two $K_1(1270)$ poles in $D^0\to \pi^+ V P$ decay,
  Phys.\ Rev.\ D {\bf 100},  074018 (2019).

\bibitem{Wang:2020pyy}
G.~Y.~Wang, L.~Roca, E.~Wang, W.~H.~Liang and E.~Oset,
Signatures of the two $K_1(1270)$ poles in $D^+\rightarrow \nu e^+ V P$ decay,
Eur. Phys. J. C \textbf{80}, 388 (2020).
  \end{thebibliography}
\end{document}